\documentclass[12pt]{amsart}
\usepackage{ifthen,verbatim}
\usepackage{floatflt,graphicx,graphics}
\usepackage{subcaption}
\usepackage{multirow}
\usepackage{setspace}
\usepackage[margin=4cm]{geometry}
\usepackage{colortbl}

 \usepackage{tikz} 
\usepackage{mathrsfs}
\usepackage{amsmath,amssymb,amsthm}
\usepackage{mathtools}
\usepackage{url}
\usepackage{rotating}

\numberwithin{equation}{section}
\nonstopmode
{\qed\bigskip}

\newcounter{alphabet}

\newcounter{minutes}
\divide\time by 60
\newcounter{hours}
\multiply\time by 60
\addtocounter{minutes}{-\time}

\begin{document}
\title[Quantitative evaluation of unsupervised clustering algorithms]
{Quantitative evaluation of unsupervised clustering algorithms for dynamic total-body PET image analysis}

\author[O. Rainio]{Oona Rainio}
\author[M. K. Jaakkola]{Maria K. Jaakkola}
\author[R. Kl\'en]{Riku Kl\'en}

\begin{abstract}
\textbf{Background.} Recently, dynamic total-body positron emission tomography (PET) imaging has become possible due to new scanner devices. While clustering algorithms have been proposed for PET analysis already earlier, there is still little research systematically evaluating these algorithms for processing of dynamic total-body PET images.\newline
\textbf{Materials and methods.} Here, we compare the performance of 15 unsupervised clustering methods, including K-means either by itself or after principal component analysis (PCA) or independent component analysis (ICA), Gaussian mixture model (GMM), fuzzy c-means (FCM), agglomerative clustering, spectral clustering, and several newer clustering algorithms, for classifying time activity curves (TACs) in dynamic PET images. We use dynamic total-body $^{15}$O-water PET images collected from 30 patients with suspected or confirmed coronary artery disease. To evaluate the clustering algorithms in a quantitative way, we use them to classify 5000 TACs from each image based on whether the curve is taken from brain, right heart ventricle, right kidney, lower right lung lobe, or urinary bladder.\newline
\textbf{Results.} According to our results, the best methods are GMM, FCM, and ICA combined with mini batch K-means, which classified the TACs with a median accuracies of 89\%, 83\%, and 81\%, respectively, in a processing time of half a second or less on average for each image.\newline
\textbf{Conclusion.} GMM, FCM, and ICA with mini batch K-means show promise for dynamic total-body PET analysis. 
\end{abstract}
\maketitle

\noindent\textbf{Author information.}\\ 
Oona Rainio$^1$, email: \texttt{ormrai@utu.fi}, ORCID: 0000-0002-7775-7656\\
Maria K. Jaakkola$^1$, email: \texttt{makija@utu.fi}, ORCID: 0000-0001-7199-0062\\
Riku Kl\'en$^1$, email: \texttt{riku.klen@utu.fi}, ORCID: 0000-0002-0982-8360\\
1: Turku PET Centre, University of Turku and Turku University Hospital, Turku, Finland\\
\textbf{Data availability.} Patient data is not available due to ethical restrictions.\\
\textbf{Code availability.} Available at \url{github.com/rklen/Clustering_algorithms_evaluated}
\\
\textbf{Conflict of interest.} On the behalf of all authors, the corresponding author states that there is no conflict of interest.\\
\textbf{Funding.} The research of the first author was funded by the Magnus Ehrnrooth Foundation and the Finnish Cultural Foundation, and the research of the second author by the Finnish Cultural Foundation.\\
\textbf{Ethical approval.} The study was approved by Ethics Committee of the Hospital District of Southwest Finland (Permission identification number: 22/1801/2022).
\\
\textbf{Informed consent.} All participants were at least 18 years of age and consented to their data to be used in this research.\\
\textbf{Acknowledgments.} The data used in the research was collected and processed in addition to the authors by Juhani Knuuti, Antti Saraste, Juha Rinne, Lauri Nummenmaa, Teemu Maaniitty, Hidehiro Iida, Vesa Oikonen, Sergey Nesterov, Jarmo Teuho, Henri Kärpijoki, Jouni Tuisku, Sarah Bär, Louhi Heli, and Reetta Siekkinen as part of the KOVERI project funded by Finnish Cardiovascular Foundation, State research funding, Finnish Cultural Foundation, and Research Council of Finland.

\section{Introduction}

Positron emission tomography (PET) is a nuclear medicine imaging method utilizing short-lived radioactive isotopes to visualize the function of the human body. During the imaging process, the patient is first given a dose of a radioactive tracer substance, this tracer then emits positrons while circulating through their body, and the PET scanner device records the collisions between the positrons and own electrons of the body based on the gamma radiation formed in them \cite{t04}. PET imaging is typically combined with either computed tomography (CT) and magnetic resonance imaging (MRI) which provide information about anatomical structures. Unlike many other imaging methods that only give static images, the results of PET imaging are also often presented as a dynamic sequence of images showing how the radioactive substance spreads around the patient's body \cite{r25}. Recently, new total-body PET scanners have enabled the full dynamic imaging of the patient instead of just some specific organ \cite{k23}.

When studying dynamic PET data, every point in the three-dimensional space has a time activity curve (TAC) that shows how the uptake of the radioactive tracer changes over time \cite{b03}. Since the amount of the tracer in an organ spikes depending when the substance first reaches the organ and then decreases according the half-life of the given isotope, the TACs differ from each other based on their location. It is known, for instance, that the tracer reaches the kidneys slower than the heart or the lungs but the tracer concentration also stays high longer in the kidneys. Understanding better these differences in the TACs can help us to detect abnormalities such as tumors, organ failure, or infection in the body.

In statistics, different clustering algorithms have been used to separate data points or vectors into smaller groups according to their shape \cite{x05}. This method has been also utilized to research dynamic PET images already for over 20 years \cite{k04,w02} but, since total-body imaging of humans has only become possible only of late, the earlier publications are limited to the research of a few organs located near each other. Furthermore, several of the existing articles also introduce only one potential method, the evaluation criteria varies over the articles, and the formed clusters sometimes are only visually inspected. During the past few decades, many new clustering algorithms have also been developed. Consequently, there is no systematic comparison which of the clustering algorithms work best on data from different organs around the human body.   

Our aim here is to find what clustering algorithms have potential for researching total-body dynamic PET data. We compare how 15 different unsupervised clustering methods perform for classifying TACs between five organs without any information about the location of the TACs in the space.
We evaluate them by computing their overall accuracy and within-cluster precision and recall, and consider the amount of variation in the results. We also take into account the processing times of the clustering algorithms.  

\section{Materials and methods}

\subsection{Software requirements}

The clustering algorithms are coded with Python (version: 3.9.9) \cite{pyt09} by using the library scikit-learn (version: 1.0.1) \cite{p11}. We also utilized automatic segmentation tool called TotalSegmentator (version: 1) \cite{w23} for CT images to obtain the ground-truth of the organ division for the clustering results. The PET and CT images and the output of TotalSegmentator were visually inspected with Carimas (version: 2.10) \cite{cri}. 

\begin{figure}[!tbp]
  \centering
  \begin{subfigure}[b]{0.17\textwidth}
  \centering
  \includegraphics[scale=0.7]{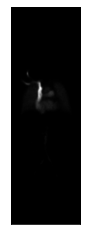}
  \caption{4 / 20s}
  \end{subfigure}
  \hspace{0cm}
  \begin{subfigure}[b]{0.17\textwidth}
  \centering
  \includegraphics[scale=0.7]{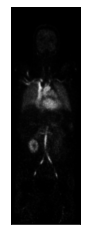}
  \caption{7 / 35s}
  \end{subfigure}
  \hspace{0cm}
  \begin{subfigure}[b]{0.17\textwidth}
  \centering
  \includegraphics[scale=0.7]{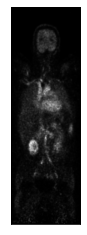}
  \caption{10 / 50s}
  \end{subfigure}
  \hspace{0cm}
  \begin{subfigure}[b]{0.17\textwidth}
  \centering
  \includegraphics[scale=0.7]{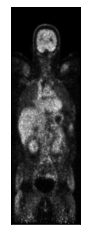}
  \caption{19 / 140s}
  \end{subfigure}
  \hspace{0cm}
  \begin{subfigure}[b]{0.17\textwidth}
  \centering
  \includegraphics[scale=0.7]{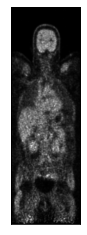}
  \caption{24 / 280s}
  \end{subfigure}
  \caption{The same coronal slice in five different time frames of a dynamic PET image of one patient. The number of the time frame and the number of seconds (s) passed since the starting time is specified in the subcaptions.}
  \label{fig1}
\end{figure}

\subsection{Data}

\begin{figure}[!tbp]
  \centering
  \begin{subfigure}[b]{0.4\textwidth}
  \centering
  \includegraphics[scale=0.45]{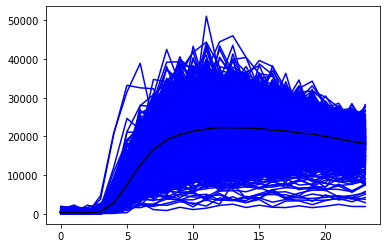}
  \caption{Brain}
  \end{subfigure}
  \hspace{0cm}
  \begin{subfigure}[b]{0.4\textwidth}
  \centering
  \includegraphics[scale=0.45]{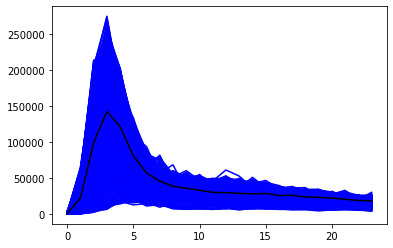}
  \caption{Right heart ventricle}
  \end{subfigure}
  \\
  \begin{subfigure}[b]{0.4\textwidth}
  \centering
  \includegraphics[scale=0.45]{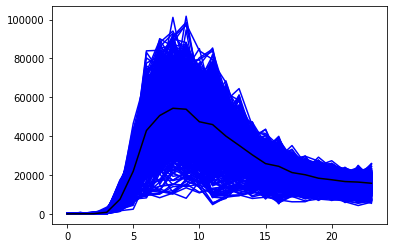}
  \caption{Right kidney}
  \end{subfigure}
  \hspace{0cm}
  \begin{subfigure}[b]{0.4\textwidth}
  \centering
  \includegraphics[scale=0.45]{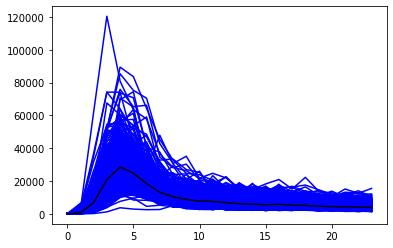}
  \caption{Lower right lung lobe}
  \end{subfigure}
  \\
  \begin{subfigure}[b]{0.4\textwidth}
  \centering
  \includegraphics[scale=0.45]{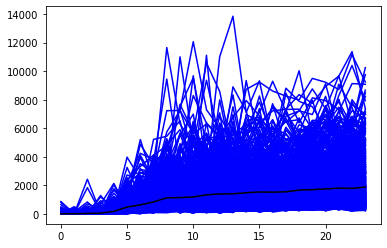}
  \caption{Urinary bladder}
  \end{subfigure}
  \caption{1000 time activity curves chosen from five different organs from a dynamic PET image of one patient. The $x$-axis varies from 0 to 23 based on the 24 time frames while the $y$-axis has the numeric elements of the TACs. The TACs themselves are in blue and their mean value curve is denoted in black.}
  \label{fig2}
\end{figure}

The data was collected from 30 patients who underwent myocardial PET perfusion imaging at Turku PET Centre in Turku, Finland, between August and November in 2022. All the participants had been either suspected to have or diagnosed coronary artery disease. The patients were injected with $^{15}$O-water and then imaged with Biograph Vision Quadra (Siemens Healthineers) PET/CT scanner. For the dynamic PET imaging, this scanner took 24 images with time intervals 14$\,\cdot\,$5s, 3$\,\cdot\,$10s, 3$\,\cdot\,$20s, and 4$\,\cdot\,$30s. See Figure \ref{fig1}. The imaging resulted in 30 CT images of $512\times512\times380$ voxels and PET images of $220\times220\times380\times24$ image points. The CT images were given to TotalSegmentator, which performed fully automatic segmentation of 104 anatomical structures, and the output was scaled into the size of $220\times220\times380$ voxels. Five different segments were chosen: brain, right heart ventricle, right kidney, lower right lung lobe, and urinary bladder. We randomly chose 1000 unique location points in each of these segments from every TotalSegmentator image and then found the corresponding TACs from the PET images, an example of which is presented in Figure \ref{fig2}. The classification task for the clustering algorithm was to separate 5000 curves of each patient based whether they are from brain, heart, kidney, lung, or bladder.

\subsection{Clustering algorithms}

We compare the following 15 methods for clustering:
\begin{itemize}
    \item The \emph{K-means} algorithm uses within-cluster sum-of-squares as a criterion for separating the samples into groups of equal variance \cite{z20}.
    \item \emph{Mini batch K-means (MBK)} is an alternative for the K-means algorithm based on using the input data in smaller subsets to reduce computation time. The MBK might produce inferior quality to K-means but this difference should be relatively small.
    \item \emph{Principal component analysis (PCA) + K-means}. PCA is a linear dimensionality reduction using singular value decomposition to find the orthogonal linear transformation for maximizing the variance of the samples \cite{g22}. By using PCA, the samples are projected on their first five principal components. After that, the K-means is used to separate the projected samples into clusters.
    \item \emph{PCA+MBK} is an alternative to the third method but with MBK instead of K-means.
    \item \emph{Independent component analysis (ICA) + K-means} is another alternative to the third method, except we use ICA instead of PCA, which means that the linear transformation is performed by using statistically independent non-Gaussian vectors as a base \cite{s02}. For ICA, too, we preserved the five first components.
    \item \emph{ICA+MBK} is an alternative to the fourth and the fifth methods with ICA instead of PCA and MBK instead of regular K-means. 
    \item \emph{Gaussian mixture model (GMM)} divides the samples into normally distributed subpopulations \cite{r09}.
    \item \emph{Agglomerative clustering (AC)} is a type of hierarchical clustering based on successively merging the samples together \cite{a14}. The linkage criteria used here is minimizing the sum of squared differences within the clusters.
    \item \emph{Spectral clustering} first performs a low-dimension embedding of the affinity matrix of the samples and then uses some algorithm, here K-means, for the clustering \cite{v07}.
    \item \emph{Birch} clustering algorithm is based on building a so-called clustering feature tree from the data and then reduces the data into a set of subclusters obtained from the tree \cite{z96}.
    \item \emph{Affinity propagation (AP)} is based on choosing a small number on such samples that best represent the others. During the clustering, AP sends messages between pairs of samples to determine how well one of them represents the other and updates its suitability criteria accordingly until convergence. AP is designed for small and medium-sized data sets. \cite{f05}
    \item \emph{Mean shift} is based on forming the clusters by choosing centroids, which are the means for the samples within a given region. Similarly to AP, the mean shift clustering finds suitable candidates by constantly adjusting its criteria \cite{c95}.
    \item \emph{Density-based spatial clustering of applications with noise (DBSCAN)} creates clusters by dividing data into high-density areas separated by low-density areas \cite{e96}.
    \item \emph{Ordering points to identify the clustering structure (OPTICS)} clustering algorithm is an alternative to DBSCAN that has more generalized criteria than a single value for defining the density necessary to form a cluster \cite{a99}.
    \item \emph{Fuzzy c-means (FCM)} is a fuzzy alternative to K-means, which first randomly divides the samples into clusters until convergence, computes the centroids of the clusters, and, for each sample and cluster, computes the degrees with which the sample belongs to the cluster \cite{z20}. After that we find the final classifications by choosing the cluster with maximal degree for every sample. 
\end{itemize}

The number of clusters was specified to be 5 for all the other methods except OPTICS, AP, mean shift, and DBSCAN, as these four algorithms do not allow this specification. All the other parameters were default settings in the functions of scikit-learn. More technical details about their implementation can be found in the scikit-learn user guide.

\subsection{Evaluation of clusters}

We processed the output of the clustering algorithms to produce five clusters: If a clustering algorithm created more than five clusters, we considered the four largest ones as they were and combined all the rests of the clusters into a single fifth cluster. If there were less than five clusters, we treated the missing clusters like they would have had zero curves. We decided which the five clusters were related to brain, heart, kidney, lung, and bladder by using the order that produces the highest accuracy. The accuracy of the results here was the percentage of curves correctly classified. For each cluster, we also computed within-cluster precision (percentage of the curves in the cluster that are correctly classified) and recall (percentage of the curves from the corresponding organ correctly classified to this cluster) \cite{r24}.

\subsection{Statistical testing}

To find out whether our methods produce statistically significant differences in their values of accuracy over all the 30 patients, we used the Wilcoxon test, which is a non-parametric alternative to the paired t-test and better suited for non-normally distributed data. The two-sided t-test was used to check whether the differences in the mean processing times were statistically significant. We also used the F-test of equality of variances to compare the variance in the accuracy between the different methods. To avoid multiple comparisons problem, the p-values were interpreted by comparing them to critical limits in the Bonferroni-corrected form $\alpha/m$, where $\alpha$ is the standard level of significance 0.05 and $m$ the number of pair-wise tests run. We used symbols * if p-value$\leq$0.05/$m$, ** if p-value$\leq$0.01/$m$, and *** if p-value$\leq$0.001/$m$ in tables.

\section{Results}

The median values of accuracy and cluster-specific precision and recall for the 15 clustering methods are collected in Table \ref{t1}. The best method in terms of overall accuracy was GMM, followed by FCM and the combination of ICA and MBK, which all had a median accuracy of over 80\%. ICA with regular K-means had a median accuracy of 78\%, which was higher than the accuracy of circa 65\% given by K-means, MBK, PCA with K-means or MBK, AC, and Birch. The rest of the methods performed very poorly: Spectral clustering, AP, DBSCAN, and OPTICS had a median accuracy of only 20\% and mean shift of 34\%. We also see that the within-cluster precision was highest for the heart and the within-cluster recall highest for the brain and the bladder.

\begin{table}[ht]
    \centering
    \begin{tabular}{|l|l|l|l|l|l|l|l|l|l|l|l|}
        \hline
        Clustering & Acc. & \multicolumn{2}{l|}{Brain} & \multicolumn{2}{l|}{Heart} & \multicolumn{2}{l|}{Kidney} & \multicolumn{2}{l|}{Lung} & \multicolumn{2}{l|}{Bladder}\\
        method & & Pre. & Rec. & Pre. & Rec. & Pre. & Rec. & Pre. & Rec. & Pre. & Rec.\\
        \hline
        K-means & 65.3 & 64.6 & \textbf{87.9} & \textbf{100.0} & 52.0 & \textbf{97.8} & \textbf{78.8} & 64.3 & 65.1 & 67.6 & \textbf{100.0}\\
        \hline
        MBK & 66.8 & 65.1 & \textbf{86.9} & \textbf{100.0} & 53.8 & \textbf{96.0} & \textbf{80.4} & 72.6 & 74.6 & 66.0 & \textbf{100.0}\\
        \hline
        PCA+K-means & 65.0 & 66.0 & \textbf{86.4} & \textbf{100.0} & 51.4 & \textbf{97.7} & \textbf{78.7} & 66.1 & 66.0 & 73.9 & \textbf{100.0}\\
        \hline
        PCA+MBK & 66.6 & 64.0 & \textbf{87.5} & \textbf{100.0} & 58.2 & \textbf{97.4} & \textbf{78.6} & 65.7 & 67.5 & \textbf{77.3} & \textbf{100.0}\\
        \hline
        ICA+K-means & \textbf{77.9} & \textbf{82.4} & \textbf{93.8} & \textbf{100.0} & 56.1 & \textbf{99.7} & \textbf{85.8} & 69.9 & \textbf{76.6} & 67.5 & \textbf{99.8}\\
        \hline
        ICA+MBK & \textbf{81.4} & \textbf{82.3} & \textbf{93.1} & \textbf{100.0} & 55.0 & \textbf{99.8} & \textbf{85.1} & 71.6 & \textbf{78.1} & 74.0 & \textbf{99.9}\\
        \hline
        GMM & \textbf{89.0} & \textbf{77.0} & \textbf{96.8} & \textbf{96.2} & \textbf{95.7} & \textbf{93.2} & \textbf{97.0} & \textbf{91.1} & \textbf{94.4} & \textbf{100.0} & \textbf{84.9}\\
        \hline
        AC & 66.0 & 56.7 & \textbf{96.0} & \textbf{99.8} & 53.4 & \textbf{99.7} & \textbf{80.6} & 56.6 & 62.7 & \textbf{84.8} & \textbf{99.7}\\
        \hline
        Spectral & 20.5 & 24.7 & 2.2 & 23.7 & 2.6 & 21.0 & 3.0 & 23.9 & 2.2 & 24.3 & 2.7\\
        \hline
        Birch & 66.0 & 56.7 & \textbf{96.0} & \textbf{99.8} & 53.4 & \textbf{99.7} & \textbf{80.6} & 56.6 & 62.7 & \textbf{84.8} & \textbf{99.7}\\
        \hline
        AP & 20.0 & 0.0 & 0.0 & 0.0 & 0.0 & 0.0 & 0.0 & 0.0 & 0.0 & 20.0 & \textbf{100.0}\\
        \hline
        Mean shift & 34.1 & 23.9 & \textbf{100.0} & \textbf{99.8} & 69.7 & 0.0 & 0.0 & 0.0 & 0.0 & 0.0 & 0.0\\
        \hline
        DBSCAN & 20.0 & 20.0 & \textbf{100.0} & 0.0 & 0.0 & 0.0 & 0.0 & 0.0 & 0.0 & 0.0 & 0.0\\
        \hline
        OPTICS & 20.8 & 25.1 & 0.8 & 71.4 & 1.6 & 45.2 & 1.0 & 73.7 & 0.9 & 20.3 & 1.0\\
        \hline
        FCM & \textbf{83.1} & 74.9 & \textbf{89.7} & \textbf{99.9} & 64.6 & \textbf{93.7} & \textbf{78.0} & \textbf{76.1} & \textbf{86.1} & \textbf{84.8} & \textbf{100.0}\\
        \hline
    \end{tabular}
    \caption{The median values of accuracy (\%) over all clusters and both the within-cluster precision (\%) and recall (\%) for the clusters of brain, heart, kidney, lung, and bladder separately computed from the results of the 15 different clustering methods for 5000 time activity curves of each of the 30 patients. The values over 75\% are in bold.}
    \label{t1}
\end{table}

The results from the Wilcoxon tests comparing the differences in the accuracy of the methods are collected in Table \ref{t2}. We see that the three best methods did not statistically differ from each other in terms of accuracy. Similarly, there were no significant differences between K-means, MBK, PCA with K-means, PCA with MBK, AC, and Birch. Nearly all the other methods significantly differed from each other, most of them with p-values of 0.001 or less.

\begin{table}[ht]
    \centering
    \begin{tabular}{|l|l|l|l|l|l|l|l|l|l|l|l|l|l|l|}
        \hline
         & \begin{sideways}FCM\end{sideways} & \begin{sideways}OPTICS\end{sideways} & \begin{sideways}DBSCAN\end{sideways} & \begin{sideways}Mean shift\end{sideways} & \begin{sideways}AP\end{sideways} & \begin{sideways}Birch\end{sideways} & \begin{sideways}Spectral\end{sideways} & \begin{sideways}AC\end{sideways} & \begin{sideways}GMM\end{sideways} & \begin{sideways}ICA+MBK\end{sideways} &
         \begin{sideways}ICA+K-means\end{sideways} &
         \begin{sideways}PCA+MBK\end{sideways} &
         \begin{sideways}PCA+K-means\end{sideways} &
         \begin{sideways}MBK\end{sideways}\\
        \hline
        K-means & *** & *** $\uparrow$ & *** $\uparrow$ & *** $\uparrow$ & *** $\uparrow$ & & *** $\uparrow$ & & * & * & & & & \\
        \hline
        MBK & *** & *** $\uparrow$ & *** $\uparrow$ & *** $\uparrow$ & *** $\uparrow$ & & *** $\uparrow$ & & *** & & * & & \\
        \cline{1-14}
        PCA+K-means & *** & *** $\uparrow$ & *** $\uparrow$ & *** $\uparrow$ & *** $\uparrow$ & & *** $\uparrow$ & & & & & \\
        \cline{1-13}
        PCA+MBK & ** & *** $\uparrow$ & *** $\uparrow$ & *** $\uparrow$ & *** $\uparrow$ & & *** $\uparrow$ & & * & & \\
        \cline{1-12}
        ICA+K-means & & *** $\uparrow$ & *** $\uparrow$ & *** $\uparrow$ & *** $\uparrow$ & & *** $\uparrow$ & & & \\
        \cline{1-11}
        ICA+MBK & & *** $\uparrow$ & *** $\uparrow$ & *** $\uparrow$ & *** $\uparrow$ & & *** $\uparrow$ & &\\
        \cline{1-10}
        GMM & & *** $\uparrow$ & *** $\uparrow$ & *** $\uparrow$ & *** $\uparrow$ & ** $\uparrow$ & *** $\uparrow$ & ** $\uparrow$\\
        \cline{1-9}
        AC & * & *** $\uparrow$ & *** $\uparrow$ & *** $\uparrow$ & *** $\uparrow$ & & *** $\uparrow$\\
        \cline{1-8}
        Spectral & *** & * & *** $\uparrow$ & *** & & ***\\
        \cline{1-7}
        Birch & * & *** $\uparrow$ & *** $\uparrow$ & *** $\uparrow$ & *** $\uparrow$\\
        \cline{1-6}
        AP & *** & & & **\\
        \cline{1-5}
        Mean shift & *** & *** $\uparrow$ & *** $\uparrow$\\
        \cline{1-4}
        DBSCAN & *** & ***\\
        \cline{1-3}
        OPTICS & ***\\
        \cline{1-2}
    \end{tabular}
    \caption{The statistically significant differences in the accuracy of different methods according to the Wilcoxon tests. Due to the Bonferroni correction, we use symbol * if p-value$\leq$0.05/$m$, ** if p-value$\leq$0.01/$m$, and *** if p-value$\leq$0.001/$m$ where $m=105$ is the number of pair-wise tests here. The upwards arrow $\uparrow$ means that the method of that row has significantly higher accuracy than the method of that column.}
    \label{t2}
\end{table}

GMM, the method with the highest median accuracy, 0.89, had also the highest standard deviation in the accuracy, 0.134. However, according to F-tests, this did not differ from the amount of variation in all the nine other methods with median accuracy of at least 0.65, which also had standard deviations of at least 0.10. The difference was significant only between the ten better performing methods and spectral clustering, AP, DBSCAN, OPTICS, or mean shift. The latter five methods had standard deviations of less than 0.08 and, in particular, the standard deviation of DBSCAN was 0 as this method always produced accuracy of exactly 0.20.

The mean performing times of the clustering methods can be seen from Table \ref{t3}. The fastest methods were K-means and MBK by themselves or after PCA or ICA, and FCM, which cluster 5000 curves less than 0.25 seconds on average. GMM and DBSCAN also took less than one second. The slowest methods were AP with mean time of 182 seconds and mean shift with mean time of 44 seconds. According to t-tests, most of the methods differ significantly from each other but the absolute differences between, for instance, FCM and GMM are relatively small.

\begin{table}[ht]
    \centering
    \begin{tabular}{|l|l|l|l|l|l|}
         \hline
        K-means & 0.207 *** $\downarrow$& ICA+MBK & 0.060+0.034 *** $\downarrow$& AP & 182.409 ***\\
        \hline
        MBK & 0.044 *** $\downarrow$& GMM & 0.597 & Mean shift & 43.809 ***\\
        \hline
        PCA+K-means & 0.028+0.166 *** $\downarrow$& AC & 1.450 ***& DBSCAN & 0.466\\
        \hline
        PCA+MBK & 0.028+0.063 *** $\downarrow$& Spectral & 5.911 ***& OPTICS & 8.596 ***\\
        \hline
        ICA+K-means & 0.060+0.182 *** $\downarrow$& Birch & 2.028 ***& FCM & 0.180 *** $\downarrow$\\
        \hline
    \end{tabular}
    \caption{The mean processing times of different clustering methods in seconds for clustering 5000 curves. The statistically significant p-values of the t-tests between the given method and GMM are denoted by * if p-value$\leq$0.05/$m$, ** if p-value$\leq$0.01/$m$, and *** if p-value$\leq$0.001/$m$ where $m=14$ is the number of pair-wise tests here. The downwards arrow $\downarrow$ means that the given method is significantly faster than GMM.}
    \label{t3}
\end{table}

We also checked the number of clusters given by the four methods that do not allow the user to specify how many clusters there should be. OPTICS produced 7-29 clusters with a mean of 15.2, AP 1-94 clusters with a mean of 14.3, and mean shift 2-179 clusters with a mean of 8.3, while DBSCAN always created only one cluster. Additionally, we observed that, for spectral clustering, the cluster sizes varied between 1-4993 curves, and the size of the smallest cluster was always at most 142 curves whereas the largest cluster was at least 4109 curves. In comparison, the cluster sizes of GMM only varied between 220-2028 curves.

\section{Discussion}

The aim of this study was to offer an initial perspective on which of the clustering algorithms might be feasible for analysis of dynamic total-body PET images. We systematically compared 15 clustering algorithms with such a task where their performance could easily be evaluated in a quantitative way. We hope that this study assists the choice of the clustering algorithms in future PET research. In the future, the most promising clustering algorithms could be used to perform image segmentation for the whole dynamic PET image in a fully automatic way, which would help the clinicians in their work. 

According to our results, the best methods are GMM, FCM, and ICA with MBK. They all had a median accuracy of over 80\% with no statistically significant differences between neither in their values of accuracy nor the amount of variation in it between the methods. These methods performed the clustering tasks of 5000 curves of 24 elements in length in 0.1-0.6 seconds on average. Interestingly, while MBK should produce slightly more inaccurate results than regular K-means according to the user manual of scikit-learn, MBK did in fact perform significantly better than K-means when combined with ICA.

K-means, MBK, PCA with K-means, PCA with MBK, ICA with regular K-means, AC, and Birch all performed decently in our experiment, resulting in median accuracies between 65-78\%. MBK did not significantly differ from K-means in terms of accuracy but the average processing time of the former method was less than one fourth of that of the latter. Unlike for ICA, the use of PCA before K-means or MBK did not improve the results significantly. It also seems that the methods based on K-means or the hierarchical type approaches of AC and Birch are more effective than the methods like AP and mean shift that aim to find somehow representative samples by comparing them with each other.

All the four methods for which the numbers of clusters could not be specified performed very poorly. One of them, DBSCAN, did not work properly because it always created only one cluster, likely because of our use of the default parameters instead of trying to find the optimal cluster density limit. Three of the other methods, OPTICS, AP, and mean shift, produced on average much more clusters than needed and might therefore work better if there were a more advanced technique for combining clusters. We did not want to choose the final clusters in a way that would maximize their accuracy here in order to keep their results comparable with the other methods. Another issue with these three methods is their slow processing times. In particularly, AP is designed for smaller data sets and, since it runs until convergence, it takes multiple times the amount of the time used by all the other methods combined.

In earlier research, Zaidi et al. \cite{z02} also noted the potential of FCM for PET processing, though for attenuation correction. On the other hand, Koivistoinen et al. \cite{k04} suggested that FCM might not separate certain brain structures. GMM has also been suggested for PET analysis by Tafro and Seršić \cite{t24}. However, Wong et al. \cite{w02} highlighted that the difficulty of determining the optimal number of clusters, which impacts both FCM and GMM.

One unexpected result was that the spectral clustering did not work properly either. Namely, it has been recommend much in the earlier research for image processing \cite{j14}. However, this method produced very unevenly sized clusters here by always placing over 80\% of the curves into a single cluster for some reason. Potentially, there could be some transformation that should be used for the data before the low-dimension embedding of spectral clustering.

Several topics could be studied further. Naturally, in later research, the most promising methods should be investigated by considering also the information about the TAC locations and different pre-processing techniques. Additionally, the impact of the potential diseases on TACs and to the clustering should be taken into account. The unsupervised methods of this article could be compared with different supervised clustering methods for processing dynamic PET images. Also, it could be studied how the information about location points affects the results and how well region growing clustering methods perform. The effect of noise in the TACs could be studied. It could be also tested how the segmentation based on the clustering algorithms works in comparison with the typical image segmentation by convolutional neural networks. Furthermore, different PET tracer substances might produce differences in the results.

\section{Conclusion}

We compared the performance of 15 clustering methods for classifying 5000 TACs from dynamic $^{15}$O-water total-body PET images of 30 patients between five different organs. According to our results, the best methods are GMM, FCM, and ICA with MBK, which give a median accuracy of over 80\% and perform the classification task from one image in a time of half a second or less. In comparison, spectral clustering, DBSCAN, OPTICS, AP, and mean shift performed here very poorly but they might work better with more deliberate choices for initial parameters and post-processing procedures. The heart has the highest precision, while the recall was highest for recognizing TACs from brain and urinary bladder.

\end{document}